\documentclass[aip,pop,10p,sanserif,reprint,onecolumn]{revtex4-1}

\usepackage[dvips]{graphicx}
\usepackage{amsmath,amsthm, amssymb, amsbsy, latexsym}
\usepackage{multirow}
\usepackage{color}

\begin{document}

\title{The ULF wave foreshock boundary: Cluster observations}

\author{Nahuel Andr\'es}
\email[E-mail: ]{nandres@iafe.uba.ar. Corresponding author: Av. Cantilo 2620 Edificio IAFE (CP 1428), CABA, Argentina. Tel.: +5411 47890179 (int. 134) ; Fax: +5411 47868114.}
\affiliation{Instituto de Astronom\'ia y F\'isica del Espacio (CONICET-UBA), Argentina}
\affiliation{Departamento de F\'isica (FCEN, UBA), Argentina}

\author{Karim Meziane}
\affiliation{Physics Department, University of New Brunswick, Fredericton, Canada}

\author{Christian Mazelle}
\affiliation{Institut de Recherche en Astrophysique et Plan\'etologie, Toulouse, France.}

\author{Cesar Bertucci}
\affiliation{Instituto de Astronom\'ia y F\'isica del Espacio (CONICET-UBA), Argentina}
\affiliation{Departamento de F\'isica (FCEN, UBA), Argentina}

\author{Daniel G\'omez}
\affiliation{Instituto de Astronom\'ia y F\'isica del Espacio (CONICET-UBA), Argentina}
\affiliation{Departamento de F\'isica (FCEN, UBA), Argentina}

\begin{abstract}
The interaction of backstreaming ions with the incoming solar wind in the upstream region of the bow shock gives rise to a number of plasma instabilities from which ultra-low frequency (ULF) waves can grow. Because of their finite growth rate, the ULF waves are spatially localized in the foreshock region. Previous studies have reported observational evidence of the existence of a ULF wave foreshock boundary, whose geometrical characteristics are very sensitive to the interplanetary magnetic field (IMF) cone angle. The statistical properties of the ULF wave foreshock boundary (UWFB) are examined in detail using Cluster data. A new identification of the UWFB is presented using a specific and accurate criterion for identification of boundary crossings. This criterion is based on the degree of IMF rotation as Cluster crosses the boundary. The obtained UWFB is compared with previous results reported in the literature as well as with theoretical predictions. Also, we examined the possible connexion between the 
foreshock boundary properties and the ion emission mechanisms at the bow shock.
\end{abstract}


\maketitle

\section{Introduction}


The upstream region magnetically connected to the bow shock is known as the foreshock. This region is populated by a small fraction of the incoming solar wind particles which are reflected at different locations of the bow shock back into the solar wind. These backstreaming particles are subjected to the solar wind's \textbf{E}$\times$\textbf{B} drift, where \textbf{E}$=-$\textbf{v}$_\text{sw}\times$\textbf{B}, is the solar wind's convective electric field, \textbf{B} is the interplanetary magnetic field (IMF) and $\textbf{v}_\text{sw}$ is the solar wind velocity. As a result, the guiding centers of all backstreaming particles move within the $\textbf{v}_\text{sw}$-\textbf{B} plane, gradually drifting away from the field line tangent to the bow shock toward the inner part of the foreshock and being segregated according to their parallel velocities. Backstreaming ions, can drive a number of plasma instabilities \citep[e.g.][]{G1993}, leading to the generation of ultra-low frequency (ULF) waves. The ion 
foreshock is then characterized not only by the presence of backstreaming ions, but also by the generation and propagation of plasma waves around the local ion cyclotron frequency.


An early study based on Vela measurements reported by \citet{A1968} showed that a fraction of ions from the solar wind are often accelerated at the Earth's bow shock and reflected into the solar wind. Using Explorer 43 data, \citet{L1974} presented evidence of backstreaming protons in the 30-100 keV range whose presence was attributed to the interaction between 3-4 keV reflected protons and Alfv\'en waves. Observations made by the dual spacecraft ISEE allowed for the identification of different types of backstreaming ion distributions at the Earth's foreshock: reflected ions (now called field-aligned beams), intermediate, and diffuse distributions \citep{G1978,P1981,T1985}. This classification of backstreaming ion populations was made on the basis of two-dimensional velocity distribution functions and energy-time spectrograms. Further results from ISEE demonstrated the existence of gyrophase-bunched and gyrotropic backstreaming ion distributions in the foreshock \citep{G1983}. These gyrating ion 
distributions are characterized by a gyro-motion around the magnetic field, i.e., a non-vanishing perpendicular bulk velocity with respect to the background magnetic field \citep{T1985,F1986,F1986b,F1994,F1995,M1997,Me2001,M2003}. \\

The field-aligned beam (FAB) distributions populate a region located just downstream from the ion foreshock boundary. This distribution originates from the quasi-perpendicular sector of the bow shock, i.e. $\theta_{Bn}>45^\circ$, where $\theta_{Bn}$ is the angle between the IMF and the local normal to the shock. As a result, the FAB region displays an inner and an outer boundary. Within this region no ULF waves are observed \citep{P1979}. These ions are characterized by a bulk motion essentially along the IMF of the order of a few $v_{sw}$ (in the plasma reference frame) and a velocity spread of a few hundreds of km s$^{-1}$ \citep{Me2013}. Basic production mechanisms for the FABs have been investigated elsewhere \citep[e.g.][]{T1983a}. \citet{K2004} using CLUSTER multi-spacecraft measurements showed that the FABs result from effective scattering in pitch angle during reflection in the shock ramp \citep[see also][]{Mi2009}. However, a comprehensive understanding of FAB production mechanisms still needs 
clarification. Downstream from the FAB region, gyrating ion distributions are often detected in association with quasi-monochromatic ULF waves with substantial amplitudes, i.e. $\delta\textbf{B}/B\sim1$ \citep{M2003}. The gyrating ion distributions could be nongyrotropic, i.e. gyrophase-bunched, or nearly gyrotropic. There are mainly two possible mechanisms for the origin of gyrating ion distributions, the waves (produced through a beam plasma instability) trap the ions and cause the phase-bunched distribution \citep{HT1985,M2000,Me2001,M2003}, or a portion of the incoming solar wind is specularly reflected at the bow shock \citep{G1983,G1982,Me2004a}. \citet{Mo2001} studied the spatial and temporal structure of FAB and gyrating ring distributions at the quasi-perpendicular bow shock with Cluster CIS data. \citet{S2000} provided a theoretical background. These authors suggest a common origin of ring and beam populations at quasi-perpendicular shocks in the form of specular reflection and immediate pitch 
angle scattering. Using a 2-D particle-in-cell (PIC) code, \citet{S2013} analyzed the ion foreshock for quasi-perpendicular conditions. Their numerical results suggest that FABs and gyrating ion distributions can be produced from different interactions with the shock front without the need of any pitch-angle scattering induced by wave-particle interaction. Finally, diffuse ion distributions spread out a shell of nearly constant radius about a mean velocity field. The bulk velocity of diffuse ions is approximately $v_{sw}$ (in the plasma reference frame) and are found even farther away from the ion foreshock boundary in quasi-parallel shock regions, i.e. $\theta_{Bn}<45^\circ$ \citep{BM1981a}. Possible mechanisms of generation of diffuse ion distributions can be found elsewhere \citep{S1987,G1989,JE1991}. Diffuse distributions are in fact observed in the presence of non-linear, steepened ULF waves \citep{P1979,P1981,H1981} and associated with quasi-parallel shock processes \citep{G1989}. \\


The most frequent type of ULF waves observed in the Earth's ion foreshock are large amplitude waves with periods from about 20 to 40 seconds, the so-called ``30 second'' waves. Their waveforms vary from quasi-monochromatic, coherent and transverse to the ambient magnetic field to highly compressional and steepened (the so-called \textit{shocklets}) \citep{H1981,LR1992b,G1995,L2002,E2005,B2005}. The quasi-monochromatic 30 second waves are usually observed with a left-handed polarization in the spacecraft frame \citep{F1969}. However, this kind of waves are intrinsically right-handed and propagate against the solar wind flow \citep{H1981}. Theoretical works has established that 30 second waves are generated via the ion-ion right-hand resonant beam instability \citep{G1991,CG1997}. Using Cluster data, \citet{M2003} were the first to firmly prove this theoretical prediction from observations. Intrinsically left-handed and sunward propagating ULF waves have also been observed in the foreshock \citep{E2003}. The excitation source of these waves is still unknown.


The region of ULF wave activity is embedded in the ion foreshock. Because of the finite growth rate of effective instabilities combined with convection, these waves reach significant amplitudes away from the source region. Therefore, for quasi-stable IMF directions, the onset of waves is spatially localized in an extended surface in the ion foreshock known as the ULF wave foreshock boundary (UWFB). For a precise identification of the UWFB, the mean magnetic field is expected to perform a very slow and monotonic rotation as the spacecraft crosses this boundary. This is a central point of the boundary morphology as the UWFB orientation depends on the IMF direction. The onset of waves in coincidence with large IMF rotations would be in conflict with the quasi-steadiness condition and therefore might lead to an incorrect identification of the boundary. An early study on the UWFB by \citet{D1976} using magnetic field data from Heos-1, shows that the boundary is strongly dependent upon the cone angle $\theta_{Bx}$, the angle between the IMF direction and the $\hat{\textbf{x}}_\text{gse}$ axis. However, the low-resolution magnetic data (48 sec) used in this work induced significant uncertainties on the boundary crossings. A decade later, using ISEE 1 data, \citet{GB1986} (GB86) confirmed \citet{D1976}'s results. Even though the time resolution of magnetic field data was better, no quantitative criteria for the identification of boundary crossings was considered by GB86. As a result, their identification of the UWFB may have included crossings with large IMF rotations. Although the authors used the ULF waves onset as a qualitative criterion for boundary crossing identification, they did not include any quantitative limit on the amplitude of the fluctuations. In another study based on magnetometer data from ISEE 1 and 2, \citet{LR1992a} examined 373 bow shock outbound crossings and recorded whether or not ULF fluctuations are present immediately upstream. They found that ULF waves are present only for $\theta_{Bn}$ less 
than $\sim~ 50^{\circ}$. Given the association of foreshock particle distributions with the ULF waves, \citet{M1998} (MD98) presented a statistical investigation of the location of the onset of intermediate and gyrating ion populations in the Earth foreshock based on the Fixed Voltage Analyzer data from ISEE 1. They found that for $\theta_{Bx}=45^\circ$, the spatial location for intermediate ions coincides with the UWFB reported by GB86. To locate the ion boundary crossings, the authors adopted a measurable criterion based on an intermediate flux level between the background level in the interplanetary medium and the higher level of ion flux events. Nevertheless, they did not establish any criterion based on the magnetic field fluctuations or rotations. A more recent study based on Cluster data, \citet{Me2004b} reported for the first time FABs and gyrating ion distributions which are observed simultaneously. The authors suggested that Cluster spacecraft might be traveling tangentially to the boundary between 
these two populations. \\

Theoretical investigations on the ULF foreshock boundary are noticeably few. To the best of our knowledge, \citet{S1988} is the only self-consistent spatio-temporal study involving the interaction between energetic protons backstreaming pre-existing and hydromagnetic waves in the Earth's ion foreshock in a theoretical frame. Using a parabolic fit to the bow shock, they found the boundary for the region of compressional waves (corresponding to different IMF orientations) using a criterion based on the compressional component of the magnetic fluctuations, $\delta|\textbf{B}|$. The authors define a theoretical wave compressional boundary where $\delta|\textbf{B}|$ becomes larger than its value in the solar wind ($1.32\times10^{-2}$ nT, according to equation (29) in the text). For a cone angle $\theta_{Bx} = 45^{\circ}$, \citet{S1988} found that the compressional boundary makes an angle $\zeta = 78^{\circ}$ with the Earth-Sun axis. \\


A comparison of the studies summarized above reveals significant discrepancies in the location of the UWFB. The lack of a common criterion for a boundary crossing certainly constitutes the main reason. The aim of the present work is to investigate the UWFB using a quantitative criterion based on magnetic field measurements with a suitable time resolution for ULF wave detection. This determination will require a stationary condition for the magnetic field direction, which will be discussed in the following Sections. A precise determination of the UWFB puts strong constraints on models involving wave-particle interactions occurring upstream of the bow shock, and at the same time it sheds light on the particle emission mechanisms at the Earth bow shock.

In-situ observations made by the flux gate magnetometer (FGM) and the Cluster Ion Spectrometer (CIS) are presented in Section \ref{obs}. Observations of ULF boundary crossings are presented in Section \ref{identification}. After, we introduce the Solar Foreshock system of coordinates in Section \ref{sfc}, the statistical results are presented in Section \ref{results}. An attempt of an interpretation of the data is presented in Section \ref{dis} followed by a short conclusions in Section \ref{con}.

\section{Data Measurements}\label{obs}

The observations used for the present paper consist of magnetic field and solar wind velocity vectors and plasma density measured by Cluster spacecraft obtained upstream from the Earth bow shock during the first three years of Cluster's orbital data, i.e. from February 2001 through December 2003. 

We have used a cadence of 5 s$^{-1}$ of magnetic field data from the flux gate magnetometer (FGM) on-board Cluster to investigate the presence of ULF waves upstream from the Earth bow shock. The particle data used in the present study are from the Cluster Ion Spectrometer (CIS) experiment, which includes a top-hat electrostatic analyzer (HIA) and a mass spectrometer (CODIF), which combines a top-hat electrostatic analyzer with a time-of-flight section to measure the major species, i.e. H$^{+}$, He$^{+}$, He$^{++}$ and O$^{+}$ over an energy range 0.02-38 keV/q. The HIA detector operates according to several modes. The velocity distributions are obtained from both instruments, which accumulate full 3-D distribution functions within one spin period (4 seconds), with an angular resolution of 22.5$^\circ\times$22.5$^\circ$. For a more extensive description of FGM and CIS/CODIF Cluster experiments see \citep{B1997a} and \citep{Re2001}, respectively. If a boundary crossing occurs when the HIA instrument is not 
operating in the solar wind mode, the plasma measurements from Cluster-CIS are not reliable. In this case, we estimated the solar wind density and velocity from plasma data taken by WIND/3DP experiment. For some events, the WIND spacecraft is located upstream as far as $\sim$250 $R_E$, and therefore for these cases a time delay between WIND and Cluster spacecraft was taken into account.

\section{Identification of UWFB crossings}\label{identification}

During Cluster excursions into the solar wind, we looked for intervals displaying patterns of ULF waves in the magnetic field components. For the determination of these crossings, we made no distinction regarding the type of magnetic field fluctuations, requiring only that the transition from (or to) the wave region would be clearly apparent. For each identified crossing, we extracted \textbf{B}, \textbf{v}$_{sw}$ and \textit{n}$_{sw}$ data from the Cluster Active Archive (CAA) and the CLWeb data base ($http://clweb.cesr.fr/$). The ULF wave foreshock boundary (UWFB) is a surface embedded in the foreshock region. By definition, the region \textit{downstream} from the UWFB is the one that displays ULF wave activity, while the region \textit{upstream} does not. For the purpose of the present study, we find that the optimal time interval to compute mean values of the quantities listed above lies between three and six wave periods at each side of the boundary. We reached this conclusion after following this procedure: for a test set of 10 crossings with particularly long and quasi-monochromatic wave activity (10 periods or more) downstream from the boundary we computed the mean values of the magnetic field components for progressively shorter time intervals. We find that for time intervals of between three and six wave periods, the mean value remains within one standard deviation from the longer average and it is therefore sufficiently long to be representative of the mean magnetic field in that region.

\begin{figure}
\centering
\includegraphics[width=.45\textwidth]{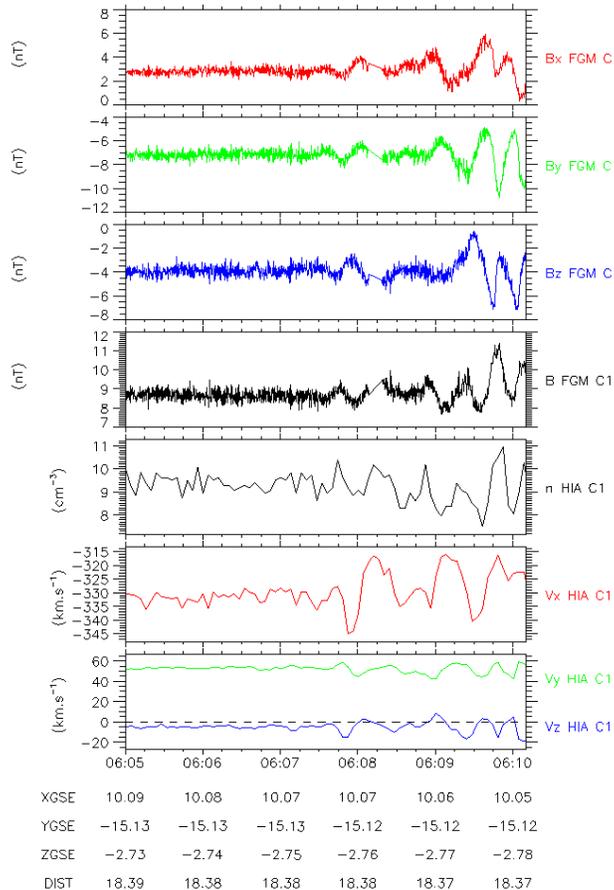}
\caption{A UWFB crossing made by Cluster on 23 April, 2001. Cluster is entering the ULF wave region with $\alpha=2^\circ\pm1^\circ$. }
\label{c1}
\end{figure}

\begin{figure}
\centering
\includegraphics[width=.45\textwidth]{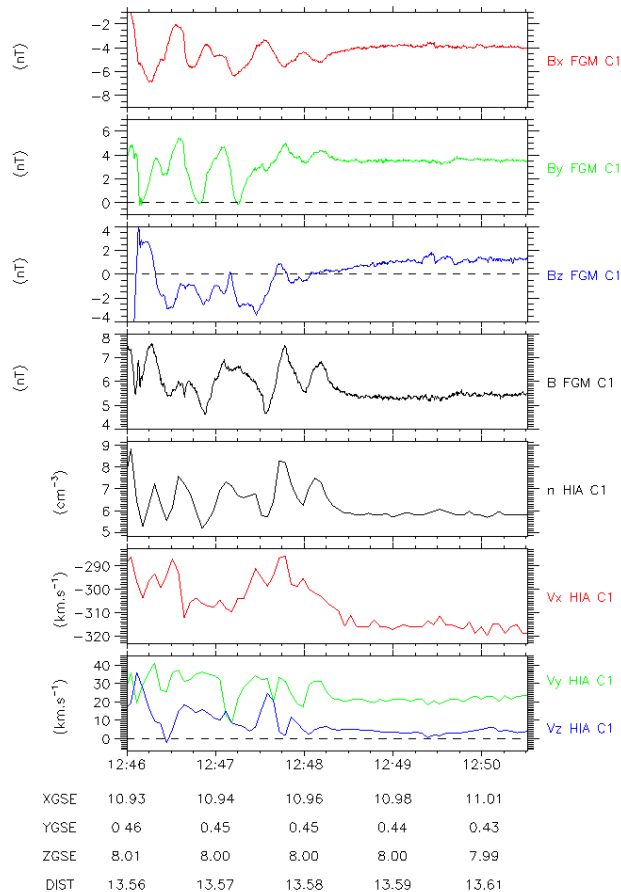}
\caption{A ULF wave region crossing (ending) made by Cluster on 16 March, 2002. In this case, $\alpha=24^\circ\pm1^\circ$.}
\label{c2}
\end{figure}

The UWFB can be observed only under quasi-stationary IMF conditions, i.e. the mean solar wind magnetic field has to remain quasi-stationary as Cluster crosses the boundary. The foreshock geometry critically depends on the IMF direction. In particular, as Cluster crosses the boundary between the upstream and the downstream region, we invariably observe a rotation of the IMF. Therefore, if we define the angle $\alpha$ as,
\begin{equation}\label{alfa}
 \cos\alpha \equiv \frac{\textbf{B}_{up}\cdot\textbf{B}_{dw}}{B_{up}B_{dw}}
\end{equation}
it is possible to quantitatively analyze the degree of IMF rotation as Cluster passes from the upstream to the downstream region. In equation \eqref{alfa}, $\textbf{B}_{up}$ ($\textbf{B}_{dw}$) is the mean magnetic field in the upstream (downstream) region, meanwhile $B_{up}$ ($B_{dw}$) corresponds to its absolute value. Then, as long as $\alpha$ remains small we are able to investigate the quasi-stationary scenario and consequently the UWFB. Figure \ref{c1} shows an example of Cluster entering to the wave region on 2001 April 23 at 0607:24 UT in which $\alpha=2^\circ\pm1^\circ$ (note that there is a small FGM data gap at 0608:15 UT in the first four panels of Figure \ref{c1}). In contrast, Figure \ref{c2} shows an example of Cluster leaving the wave region on 2002 March 16 at 1248:17 UT with $\alpha=24^\circ\pm1^\circ$.

\begin{figure}
\centering
\includegraphics[width=0.6\linewidth]{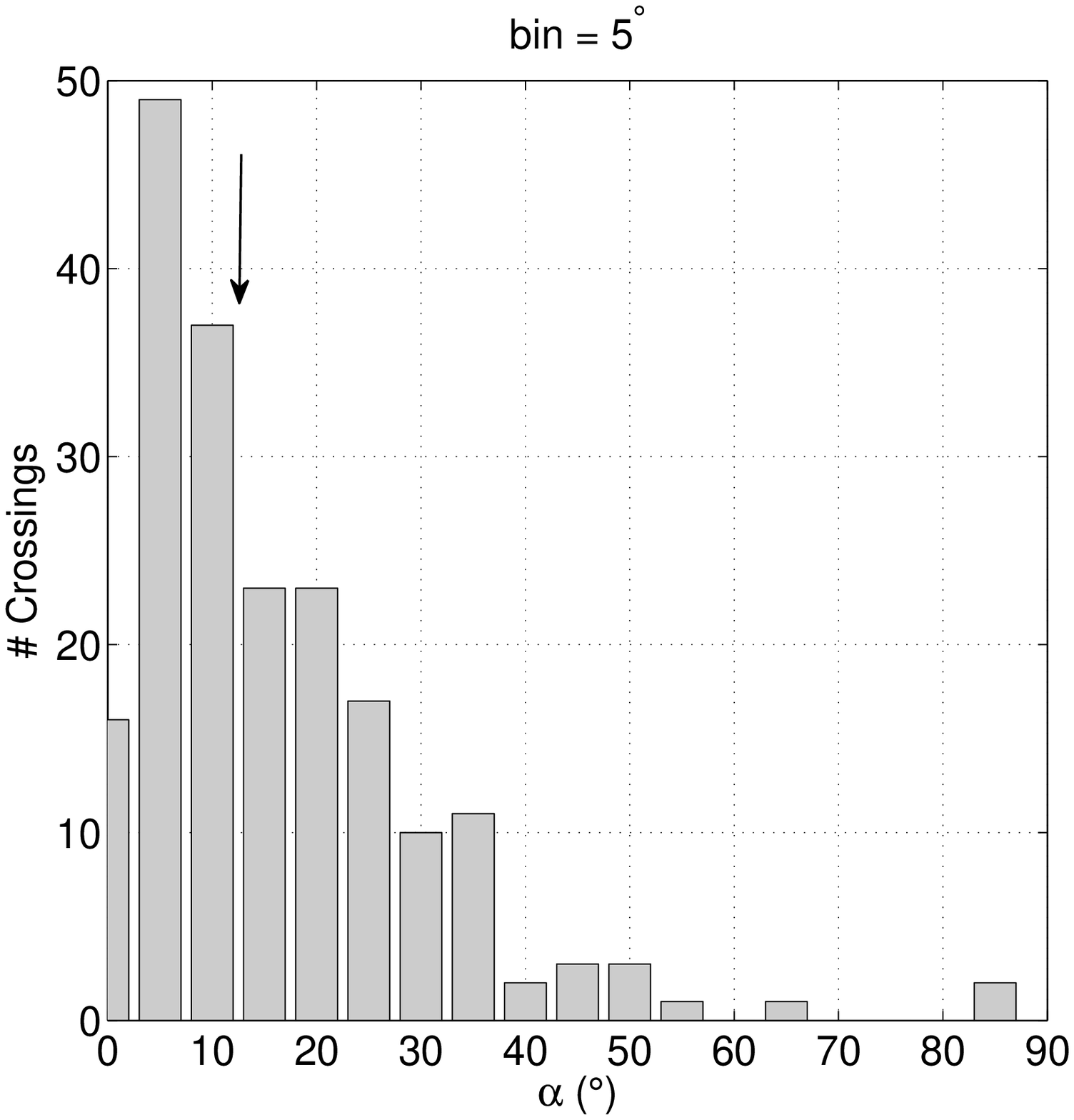}
\caption{Histogram of $\alpha$ for the 192 identified crossings. The arrow is located at $\alpha=12.5^\circ$, which is the upper limit adopted for the stationary UWFB.}\label{alpha}
\end{figure}

\begin{figure}
\centering
\includegraphics[width=.6\textwidth]{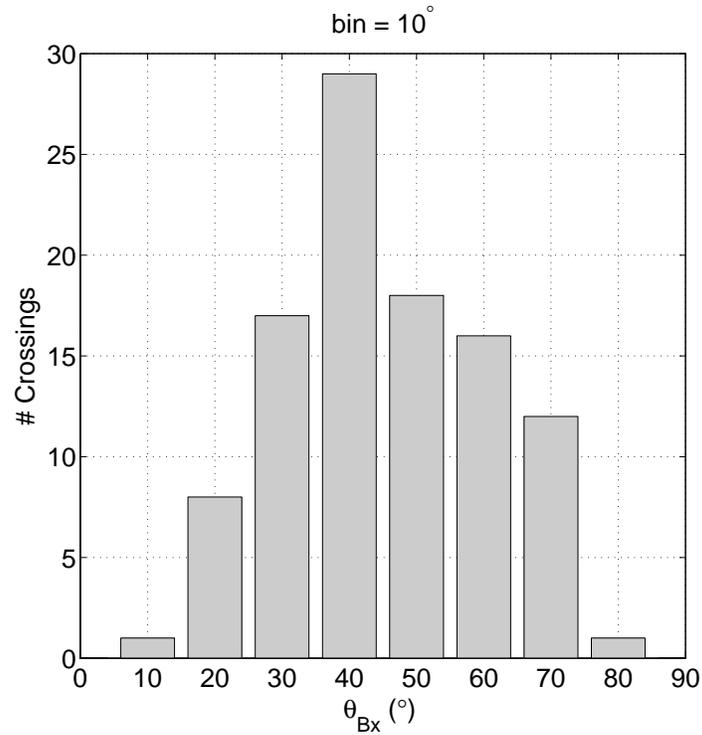}
\caption{Histogram of the cone angle $\theta_{Bx}$ for the 102 identified crossings with $\alpha<12.5^\circ$ correspond to the upstream region.}
\label{thBx}
\end{figure}

From February 2001 through December 2003, we have identified 192 ULF wave/no wave transitions. Figure \ref{alpha} shows a distribution of these transitions with respect to the $\alpha$ angle. As it can be seen, the histogram displays a strong peak at small values of $\alpha$, in particular for the most frequently value at $5^\circ$. The histogram corresponds to a non-symmetric and positively skewed distribution. Fifty percent of the crossings occur for $\alpha<12^\circ$ (i.e. $12^\circ$ corresponds to the median value of the distribution) and 78$\%$ of the events occur for angles less than the mean value of $16^\circ$. For a better determination of the boundary location, only wave crossings with $\alpha<12.5^\circ$ (black arrow in Figure \ref{alpha}) were considered, which includes 102 events. As we discussed in the Introduction, the UWFB is defined for a given IMF orientation. Figure \ref{thBx} shows the cone angle  ($\theta_{Bx}$) distribution corresponding to these 102 events. There is no indication that boundary crossings occur for a particular IMF direction other than the Parker's prediction, since the distribution is consistent with the IMF spiral orientation at 1 AU. Following previous studies, results of our statistical survey is performed using $10^\circ$ bins for the $\theta_{Bx}$.

In a previous study, \citep{A2013} presented a different approach to the determination of the UWFB crossings at Saturn. The authors considered a quasi-stationary crossing according to the following criterion: for each component of the magnetic field ($j=x,y,z$) if the difference between the average values in the upstream region (${B}_{up, \text{j}}$) and in the downstream region (${B}_{dw, \text{j}}$) is smaller than the standard deviation in the upstream region ($\sigma_{dw, \text{j}}$), they consider that the spacecraft crossed a quasi-stationary UWFB, 
\begin{equation}
|{B}_{up, \text{j}}-{B}_{dw, \text{j}}|<\sigma_{dw, \text{j}}\quad\text{for}\quad j=x,y,z
\end{equation} 
Following the same criterion established by \citet{A2013}, we found that 127 of the 192 crossings correspond to quasi-stationary crossings. In this sense, we conclude that the current criterion based on the IMF rotation is more restrictive than the criterion used by \citet{A2013}.

For our analysis, the criterion based on the magnitude ($\alpha < 12.5^\circ$) level of IMF rotation is used to identify a UWFB crossing. The determination of the UWFB spatial location requires to know the position of the Earth's bow shock. For this purpose, we use the \citet{F1991} (F91) gas dynamics bow shock model.

\section{Solar Foreshock Coordinates}\label{sfc}

In order to identify the UWFB independently from the changes in the IMF or the location of the bow shock, we employed the so-called solar foreshock coordinates (SFC) introduced by GB86. First, we construct the foreshock geometry based on the F91 bow shock model. In this model, the  bow shock location and shape are defined by the solar wind ram pressure. The bow shock model is axially symmetric about the Earth-Sun direction, and has the following functional form,
\begin{equation}\label{conic}
 r=\frac{L}{1+e\text{cos}\theta}
\end{equation}
where $r$ is the distance from the planet to a point on the shock surface, $\theta$ is the corresponding polar coordinate angle with respect to the symmetry axis, $L$ is the semilatus rectum (size parameter) and $e$ is the eccentricity. To rescale the size parameter $L$, we used the fact that the location of the bow shock varies as the inverse one-sixth power of the dynamic pressure \citep{B1968},
\begin{equation}\label{parameter}
\begin{aligned}
 L&=\bigg(\frac{p}{p_0}\bigg)^{-\frac{1}{6}}L_0 
\end{aligned}
\end{equation}
where $p$ is the ram pressure and $p_0$ is a reference ram pressure. The F91 model uses a fixed eccentricity $e=0.81\pm0.02$ (ellipsoidal model), a nominal parameter size $L_0=(24.8\pm0.2)~R_E$ with a reference ram pressure $p_0=1.8$ nPa and a nominal zero focus position. In contrast to hyperbolic models, the F91 elliptic model has not a flaring angle. F91 provides a statistical model for the shape of the Earth's bow shock. There are numerous bow shock models available in the literature and their reliability is parameter-dependent \citep{Mer2005}. Our choice \citep{F1991} is dictated by its simplicity since it is only solar wind ram-pressure-dependent. It is clear from equations \eqref{parameter} that in-situ solar wind density and velocity measurements are necessary to accurately determine the location and shape of the bow shock.

\begin{figure}
\centering
\includegraphics[width=.7\textwidth]{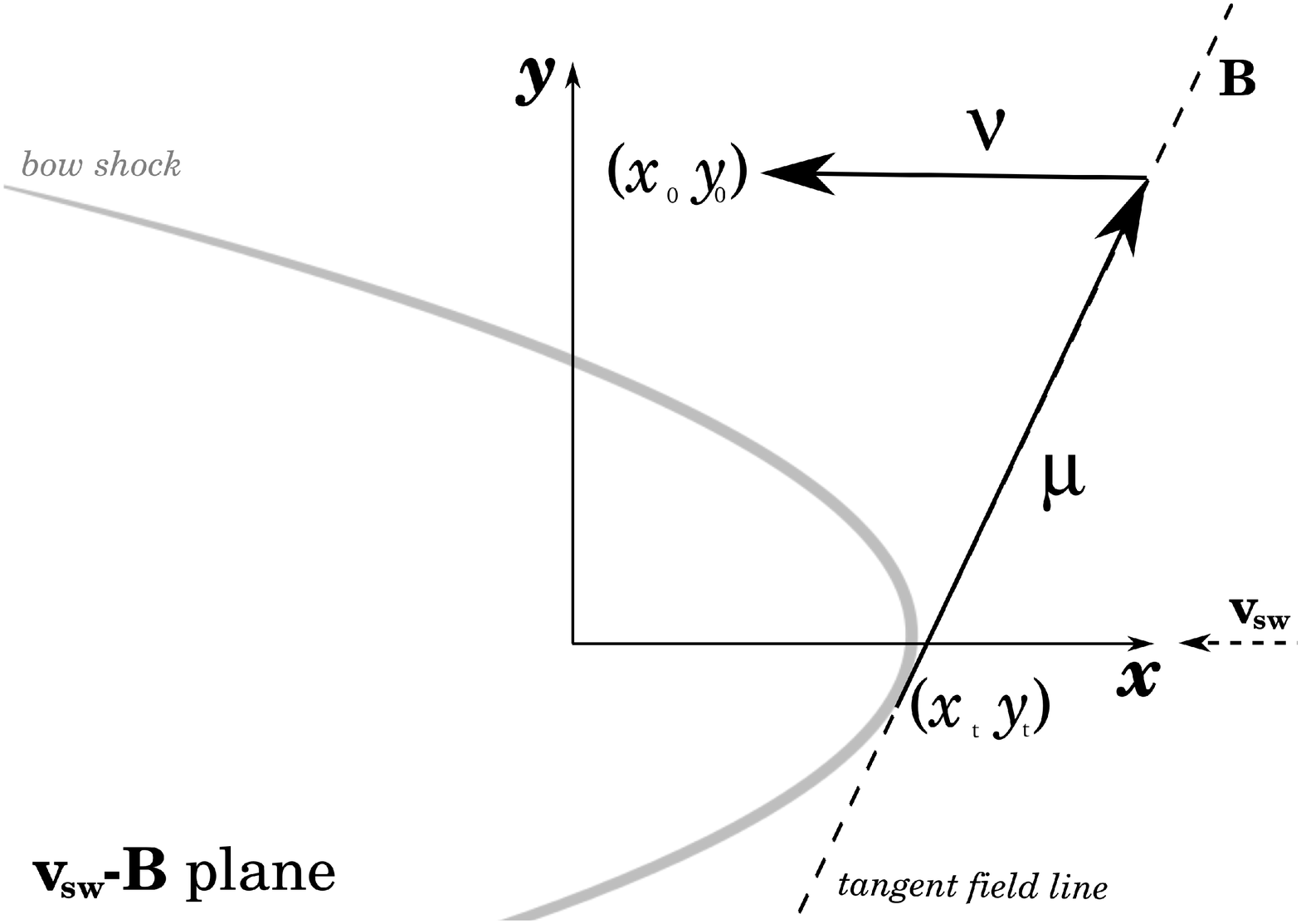}
\caption{Schematic representation of the Solar Foreshock Coordinates (SFC).}
\label{gb1}
\end{figure}

In the SFC coordinate system, the {$\hat{\textbf{x}}$} axis points toward the Sun (it is parallel to $\hat{\textbf{x}}_\text{gse}$), and the {$\hat{\textbf{x}}$}-{$\hat{\textbf{y}}$} plane is the $\textbf{v}_{\text{sw}}$-$\textbf{B}$ plane which contains the location of the spacecraft at a given quasi-stationary crossing of the UWFB. The position of each boundary crossing is fixed by calculating the SFC coordinates $\mu$ and $\nu$ (see Figure \ref{gb1}).
\begin{equation}\label{SFC}
\begin{aligned}
\mu &= \frac{(y_o-y_{t})}{\sin\theta_{\text{Bx}}} \\
\nu &= \frac{(y_o-y_{t})}{\tan\theta_{\text{Bx}}} + x_{t} -x_o
\end{aligned}
\end{equation}
where $\theta_{Bx}$ is the IMF cone angle, ($x_t,y_t$) and ($x_o,y_o$) are the IMF line tangent point to the bow shock model and the observation point respectively. The coordinate $\mu$ is the distance along the tangent magnetic field line between the tangent point and the observation point. The coordinate $\nu$ is the distance along the $\hat{\textbf{x}}_\text{gse}$ direction between the tangent magnetic field line and the observation point. In this sense, the coordinate $\nu$ indicates how far downstream from the IMF tangent line is the UWFB.

\section{Statistical Results}\label{results}

\begin{figure}
\centering
\includegraphics[width=.5\textwidth]{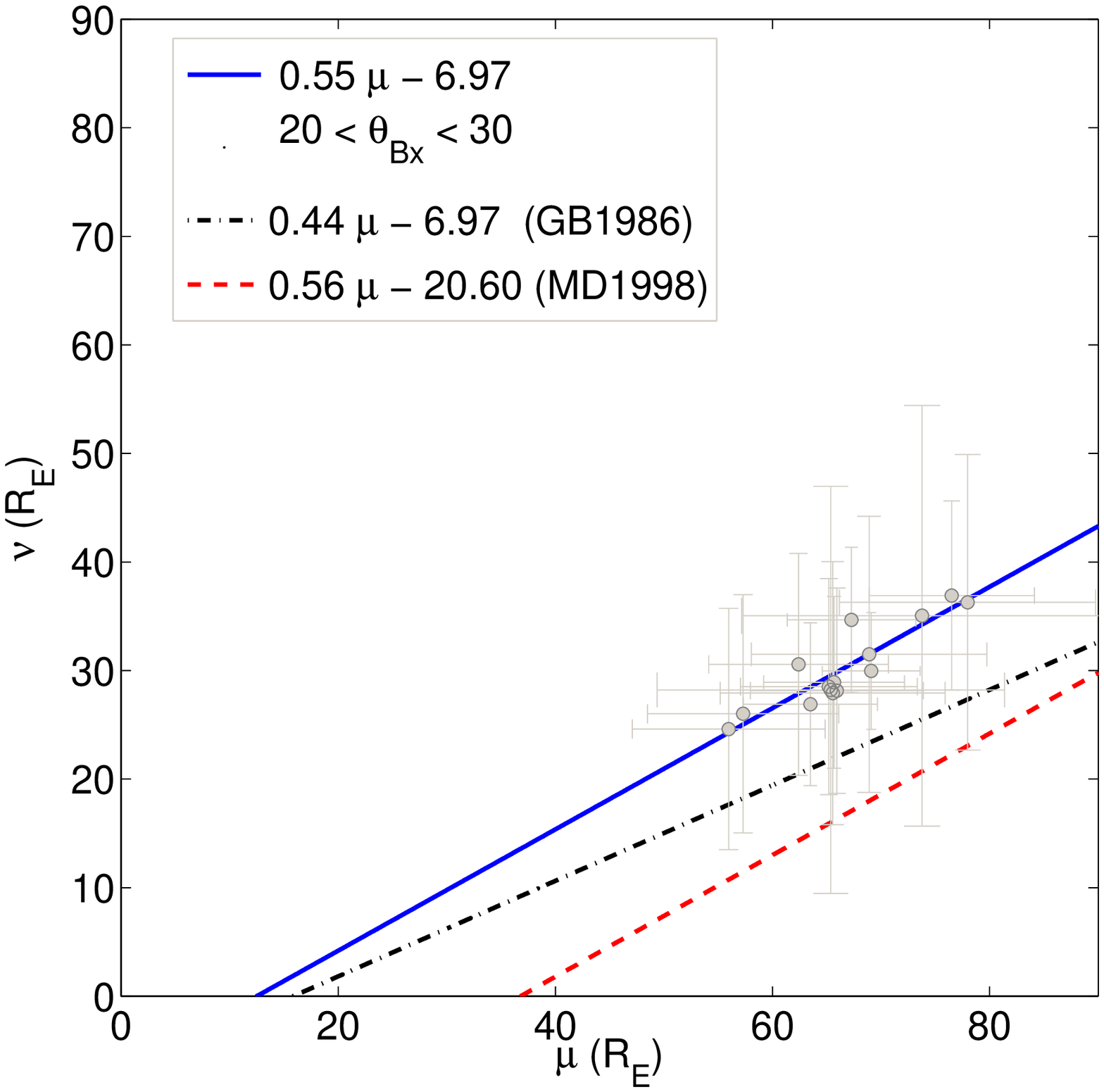}\includegraphics[width=.5\textwidth]{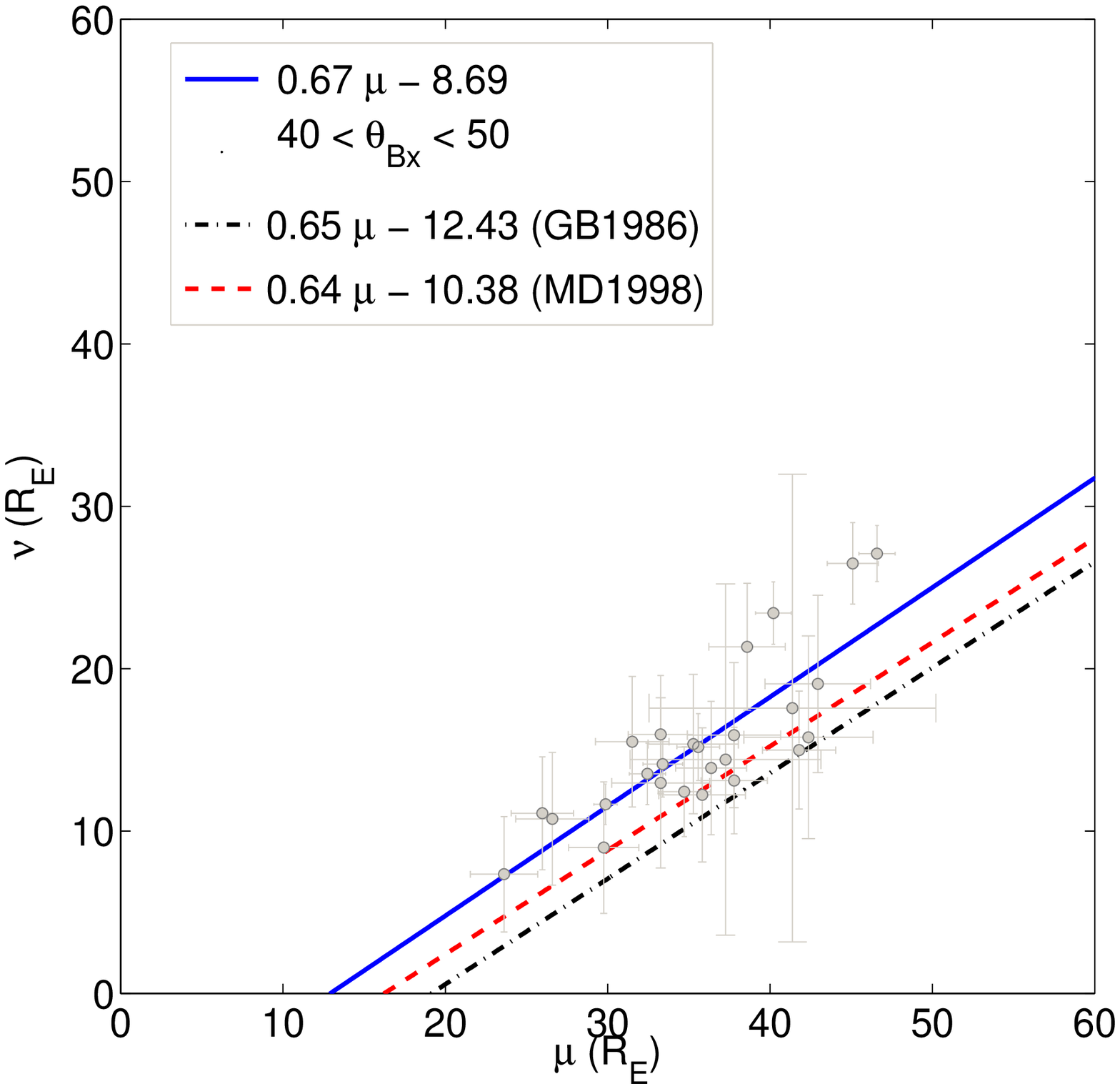}
\caption{Best linear fit (solid line) of UWFB crossings in SFC, for cone angles $40^\circ<\theta_\text{Bx}<50^\circ$ and $20^\circ<\theta_\text{Bx}<30^\circ$. For reference, we included MD98 (dashed line) and GB86 (dot-dashed line) results.}
\label{figsfc}
\end{figure}

Figure \ref{figsfc} shows a scatter plot of the UWFB crossings in SFC for two cases: $20^\circ<\theta_\text{Bx}<30^\circ$ (in the left panel) and $40^\circ<\theta_\text{Bx}<50^\circ$ (in the right panel). In these Figures, the best linear fit corresponds to the blue solid line. The uncertainties in the determination of the three magnetic field components, the three solar wind velocity components and the solar wind density are dominated by the statistical error given by the corresponding standard deviation from the mean values. For the determination of the error bars in Figure 6, we propagated these statistical errors. For reference, we included MD98 (dashed red line) and GB86 (dot-dashed line) results. It is worth mentioning the very good agreement between our results and those reported in previous works.

Table \ref{1} shows the parameters of the $\mu$-$\nu$ regression line ($\nu = p\mu + q$) of the UWFB for different $\theta_{Bx}$ ranges, using the F91 bow shock model. For comparison, the results reported by MD98 and GB86 also indicated in the third and fourth column. Except for the range $30^\circ<\theta_\text{Bx}<40^\circ$, our determination of the slope of the UWFB line is in good agreement with those obtained by MD98 (when the errors are taken into account). On the other hand, our results for the range $20^\circ<\theta_\text{Bx}<30^\circ$ differ significantly from those reported by GB86. Also, based on numerical values listed in Table \ref{1}, an increase of the UWFB line slope with $\theta_{Bx}$ in the SFC plane cannot be ruled out. However, the values of $q$ reported in MD98 are in general larger than those obtained in the present study. We suspect that the difference is due to the different bow shock models used in these studies. MD98 used a hyperboloid shape with constant parameters, while in the present work the parameters of the elliptical bow shock are adjusted according to plasma data for each boundary crossing.

\begin{table}\normalsize
\centering
\caption{Parameters of the $\mu$-$\nu$ regression line of the UWFB for different $\theta_{Bx}$ ranges, using F91 bow shock model. For comparison, we show the coefficients reported by MD98 and GB86.}
\begin{tabular}{c|c c c|c c|c c}
\hline
$\theta_{Bx}$ ($^\circ$) & $p$ & $q$ & $r$ & $p_{MD}$ & $q_{MD}$ & $p_{GB}$ & $q_{GB}$\\
\hline
20-30 & 0.56 $\pm$ 0.08 & -6-97 $\pm$ 5.21  & 0.8937 & 0.56 $\pm$ 0.02 & -20.60 $\pm$ 2.45 & 0.44 $\pm$ 0.03 & -6.97 $\pm$ 3.88\\
30-40 & 0.55 $\pm$ 0.08 & -6.43 $\pm$ 3.66  & 0.8510 & 0.71 $\pm$ 0.03 & -20.22 $\pm$ 1.48 & - & -\\
40-50 & 0.67 $\pm$ 0.09 & -8.69 $\pm$ 3.48 & 0.8192 & 0.64 $\pm$ 0.02 & -10.38 $\pm$ 0.68 & 0.65  $\pm$ 0.04 & -12.43 $\pm$ 1.56\\
50-60 & 0.77 $\pm$ 0.10 & -10.40 $\pm$ 3.14 & 0.8737 & 0.66 $\pm$ 0.02 & -9.23  $\pm$ 0.61 & - & -\\
60-70 & 0.65 $\pm$ 0.24 & -5.54 $\pm$ 6.05 & 0.6278 & 0.67 $\pm$ 0.03 & -8.48  $\pm$ 0.64 & - & - \\
\hline
\end{tabular}\label{1}
\end{table}

\section{Physical Implications}\label{dis}

One of the most interesting features of the ion foreshock is the interaction of backstreaming ions with the incoming solar wind is the excitation of ULF waves \citep{G1993}. Near the leading edge of the ion foreshock boundary, FABs collimated along IMF lines have been observed upstream from the quasi-perpendicular shocks without the presence of ULF waves \citep{P1979}. Deeper into the foreshock region, intermediate, gyrating and diffuse distributions are usually observed in association with large amplitude ULF waves \citep{P1981,H1981}. In particular, occurrence of these ULF waves are associated with gyrating ion distributions, while FABs are observed just adjacent to the interval of wave occurrence. Recently, \citet{Me2004b} reported for the first time a local energetic ion event presenting a clear double-peak spectrum observed by Cluster at approximately $1 R_E$ upstream from Earth's bow shock. The lower energy peak was associated with FAB distributions with a steady IMF, while the higher energy peak was interpreted as gyrating ion distributions having pitch angles of about $30^\circ$ in association with quasi-monochromatic ULF waves. The authors interpreted the simultaneous observation of the two distinct populations as a UWFB crossing.

In this context and in agreement with \citet{S1988}, \citet{Me2004b} reported a boundary between the FAB and the gyrating regions, which forms an angle of $\zeta=77^\circ \pm 3^\circ$ with respect to the Earth-Sun axis for a single event. As mentioned in the Introduction, during times when the FABs were observed, the mean magnetic field was steady, while the appearance of the gyrating ion distribution was accompanied by the presence of ULF waves. The authors claim that the changes observed in the proton distributions are likely due to a very weak IMF rotation. However, they did not report explicitly the level of IMF rotation associated with this crossing. \citet{S1988}'s compressive criterion is different from the two criteria discussed in this paper (see Section \ref{identification}). Whereas the criterion used in the present study is based on the magnitude of the IMF rotation, \citet{S1988} solve for locations where the compressive wave amplitude $\delta|\textbf{B}|$ rises above its ambient solar wind 
value. Furthermore, the authors claim that the observed boundary would be farther downstream, where they expect the wave amplitudes to be larger. In the case of $\theta_{Bx}=45^\circ$, the compressional boundary has a mean inclination of 78$^\circ$ with respect to the Earth-Sun axis which is significantly less than the observed value reported here  ($\zeta=87^\circ \pm 7^\circ$). The difference may be explained by the fact that the observed boundary is based on the wave onset having a larger amplitude than used by \citet{S1988} and therefore the UWFB is expected to be located downstream. Moreover, our results clearly indicate that the slope of the UWFB with respect to the $x$-axis increases with $\theta_{Bx}$ angle, which is consistent with the foreshock global structure. \\

The characterization of the UWFB is an important way of testing the validity of models considering local wave-particle interactions in the foreshock region. The spatial distribution and the dynamics of backstreaming ion distributions in the Earth foreshock has been extensively studied in the literature \citep{G1978,P1981,BM1981a,BM1981b,HR1983}. Based on the ratio $P_{gc} = v_{gc}/v_{sw}$ ($v_{gc}$ is the guiding center velocity of backstreaming ions measured in the spacecraft frame and $v_{sw}$ is the solar wind velocity), \citet{BM1981a} determined statistical average values $<P_{gc}>$ for different backstreaming ion distributions regardless of the IMF cone angle $\theta_{Bx}$. In particular, the authors find that $<P_{gc}> = 2$ for FAB distributions, $<P_{gc}> = 1.75$ for intermediate distributions, and $<P_{gc}> = 1.18$ for diffuse distributions. From the histograms reported in the paper \citep{BM1981a} we estimated a spread of $\pm 0.5$ for each $<P_{gc}>$. Considering  a fictitious proton beam propagating along the UWFB, MD98 used a straightforward geometric argument to demonstrate that the value of $P_{gc}$ can be, for a fixed $\theta_{Bx}$, related to the boundary parameters. The value of $P_{gc}$ could be interpreted as the bulk ion velocity of ions traveling along the UWFB (normalized to the solar wind speed). Using the same expressions from MD98, as indicated on the caption of Table \ref{2}, we found that for $\theta_{Bx}=45^\circ$, the $P_{gc}$ value associated with the UWFB is $1.05 \pm 0.01$, in approximate agreement with MD98 findings ($P_{gc} = 1.10\pm0.04$). The numerical value of $P_{gc}$ is therefore consistent with the bulk speed of diffuse ions rather than intermediate or gyrating ion distributions. The obtained numerical value for P$_{gc}$ may seem puzzling if we assume the the UWFB corresponds to the waves onset, where one expects the presence of an early phase of a FAB disruption. We emphasize that the large spread ($\pm 0.5$) in the determination of a particular ion distribution by \citet{BM1981a} stems from the fact that they do not consider the IMF cone angle in their classification, which could lead to misinterpret our results. However, this result strongly contrasts with the numerical value associated with the FAB-gyrating boundary $P_{gc} = 1.68\pm0.08$ \citep{Me2004b}. Our results also indicate, for $50^\circ \leq \theta_{Bx} \leq 60^\circ$, that the UWFB characteristics seems to be consistent with FAB disruption as wave excitation source, in agreement with the MD98 study. \\

An interesting aspect of the UWFB is the position with respect to the bow shock, best illustrated with the angles $\theta_{Bn}$ and $\theta_{Vn}$. Acceleration models bear a direct relation with the shock geometry. A study based on 373 bow shock crossings from \citet{LR1992a} showed that no ULF waves are present for $\theta_{Bn} \geq 50^{\circ}$. The straight-lines obtained from the best fit of the UWFB are not strictly intercepting the shock. However, given the uncertainties on $p$ and $q$ we were able to construct tangent lines by a rather small translation parallel to the $x$-direction. The obtained values for $\theta_{Bn}$ and $\theta_{Vn}$ along with their uncertainties are listed on Table \ref{2} for each range in $\theta_{Bx}$. If the uncertainties are taken into account, Table \ref{2} indicates that the UWFB intersects the shock at the transition between quasi-perpendicular and quasi-parallel shock regimes, i.e $\theta_{Bn}\sim45^{\circ}$. For $\theta_{Bx} \geq 40^{\circ}$, our results are in agreement with \citet{LR1992a} findings. Also, we find that a decrease of $\theta_{Bn}$ with increasing $\theta_{Bx}$ can not ruled out. \\

\begin{table}\normalsize
\centering
\caption{The $\zeta=\tan^{-1}(\frac{\sin\theta_{Bx}}{\cos\theta_{Bx}-p})$ angle between the UWFB and the $\hat{\textbf{x}}_{\text{gse}}$, the $P=\frac{\tan\zeta}{\tan\zeta\cos\theta_{Bx}-\sin\theta_{Bx}}$ value, the $P_{gc}=\sqrt{1+P^2-2P\cos\theta_{Bx}}$ factor, the angles $\theta_{Bn}$ and $\theta_{Vn}$ and the shock normalized velocity $P_{s}=\frac{\cos\theta_{Vn}}{\cos\theta_{Bn}}$.}
\begin{tabular}{c|c c c|c c c}
\hline
$\theta_{Bx}$ ($^\circ$) & $\zeta$ ($^\circ$) & $P$ & $P_{gc}$ & $\theta_{Bn}$ ($^\circ$) & $\theta_{Vn}$ ($^\circ$) & P$_{s}$ \\
\hline
20-30 & 50.66 $\pm$ 0.29 & 1.78 $\pm$ 0.49 & 0.97 $\pm$ 0.01 & 53 $\pm$ 5  & 28 $\pm$ 6 & 1.46 $\pm$ 0.18  \\
30-40 & 64.86 $\pm$ 0.64 & 1.81 $\pm$ 0.41 & 1.15 $\pm$ 0.01 & 52 $\pm$ 3  & 18 $\pm$ 3 & 1.54 $\pm$ 0.10  \\
40-50 & 86.99 $\pm$ 6.00 & 1.49 $\pm$ 0.21 & 1.05 $\pm$ 0.01 & 47 $\pm$ 4  & 4 $\pm$ 3 & 1.46 $\pm$ 0.10   \\
50-60 & 103.50 $\pm$ 2.13 & 1.29 $\pm$ 0.88 & 1.09 $\pm$ 0.01 & 42 $\pm$ 3  & 14 $\pm$ 3 & 1.30 $\pm$ 0.06  \\
60-70 & 104.10 $\pm$ 4.20 & 1.53 $\pm$ 3.50 & 1.43 $\pm$ 0.01 & 38 $\pm$ 7  & 27 $\pm$ 6 & 1.13 $\pm$ 0.12  \\
\hline
\end{tabular}\label{2}
\end{table}

Following MD98, we now examine the possible connexion between the UWFB properties and the ion emission mechanisms at the shock. Since the UWFB results from waves excitation generated be shock-accelerated ion beams, a resonant interaction requires a specific beam speed. Briefly, the main shock emission mechanisms are:
\begin{enumerate}
\item Magnetosheath particle leakage: \citet{E1982} studied the presence of upstream distributions which have leaked from the magnetosheath conserving their magnetic moment. In order to escape to the upstream region, these magnetosheath particles must at least reach the speed of the bow shock, which can be regarded as a threshold. In particular, this process explains the observation of low energy ion beams \citep{E1982,Ta1983,T1983b}. The predicted normalized velocity is,
\begin{equation}
P_{m.l.} = P_{s} = \frac{\cos(\theta_{Vn})}{\cos(\theta_{Bn})}
\end{equation}
where $P_{s}$ is the shock velocity or the deHoffman-Teller velocity in the plasma frame of reference normalized to $v_{sw}$.
\item Adiabatic reflection: a portion of the solar ions produces an ion beam aligned with the IMF with generally high energies \citep{S1969,T1983b}. The reflected ions acquire a speed in the plasma reference frame (normalized to $v_{sw}$) given by,
\begin{equation}\label{adiabatic}
P_{a.r.} = 2P_{s}
\end{equation}
\item Specular reflection of a portion of the solar wind ions gives birth to an emission mechanism in the upstream region only when $\theta_{Bn}<45^\circ$ \citep{S1983}. Incident solar wind ions simply reverse their component of velocity parallel to the shock normal \citep{T1983b}. In this case the post-encounter parallel velocity in the plasma reference frame  (normalized to $v_{sw}$) is given by,
\begin{equation}
P_{s.r.} = 2\cos(\theta_{Vn})\cos(\theta_{Bn})
\end{equation}
Table \ref{3} summarizes the numerical values for each emission mechanism for each interval in $\theta_{Bx}$.
\end{enumerate}

\begin{table}\normalsize
\centering
\caption{Predicted velocities (normalized to $v_{sw}$) for the main shock emission mechanisms, i.e. magnetosheath particle leakage ($P_{m.l.}$), adiabatic reflection ($P_{a.r.}$) and specular reflection ($P_{s.r.}$).}\label{sampletable}
\begin{tabular}{c|c c c}
\hline
$\theta_{Bx}$ ($^\circ$) & $P_{m.l.}$  & $P_{a.r.}$ & $P_{s.r.}$   \\
\hline
20-30 & 1.46 $\pm$ 0.18 & 2.92 $\pm$ 0.36 & 1.06 $\pm$ 0.13 \\
30-40 & 1.54 $\pm$ 0.10 & 3.08 $\pm$ 0.20 & 1.17 $\pm$ 0.08 \\
40-50 & 1.46 $\pm$ 0.10 & 2.92 $\pm$ 0.20 & 1.36 $\pm$ 0.10 \\
50-60 & 1.30 $\pm$ 0.06 & 2.60 $\pm$ 0.12 & 1.44 $\pm$ 0.07 \\
60-70 & 1.13 $\pm$ 0.12 & 2.26 $\pm$ 0.24 & 1.40 $\pm$ 0.15 \\
\hline
\end{tabular}
\label{3}
\end{table}
The third column in Table \ref{2} lists the numerical values of the normalized velocity $P$ in the rest frame of an ion propagating along the UWFB. $P$ is directly related to the slope of the boundary \citep{M1998}. This last value is compared to the normalized shock speed $P_{s}$ given in the last column of Table \ref{2}. Table \ref{3} clearly indicates that the ions propagating along the boundary are fast enough to escape upstream ($P \geq P_s$ for all $\theta_{Bx}$ values). The numerical values of $P$ and $P_s$ also indicate that the observations are in very good agreement with the magnetosheath leakage model only for $\theta_{Bx}=45^\circ$ and $\theta_{Bx} = 55^\circ$ cases. On the other hand, the specular reflection model seems to be a weak model. This strongly suggests that backstreaming gyrating ion distributions resulting from specular reflection are not likely a source for wave excitation, a result which is consistent with previous studies \citep{M2003}. \citet{K2004} reported strong observational 
evidence that magnetosheath leakage is not a source mechanism for FAB production. To overcome the problem of the injection, \citet{K2004} proposed that FABs may result from pitch angle scattering of specularly reflected ions (which could occur for all shock geometries). The ions with high parallel speed would escape upstream and the resulting particle distribution would then appear as FABs. The obtained parallel speed by \citet{K2004} (which results after the pitch angle scattering) is equal to the one obtained directly from a direct reflection with conservation of the magnetic moment. Therefore, equation (7) used in the present paper could correspond to either an adiabatic reflection or to the mechanism described in \citet{K2004}. However, the boundary derived velocity is underestimated compared to the one obtained from the adiabatic reflection hypothesis.

\section{Conclusions}\label{con}

Using a well defined, accurate and robust criterion we present a new determination of the boundary for ULF waves in the Earth foreshock (which we termed UWFB). Our criterion allows to quantitatively measure differences between the magnetic field upstream and downstream from the UWFB, taking into account possible rotations of the IMF. \\

All the wave events reported in the present paper show evidence of magnetic connectivity to the Earth bow shock models, which is a clear indication that these waves are associated with the Earth foreshock. As mentioned in Section \ref{identification}, the precise location of the UWFB is only determined under quasi-stationary IMF conditions. The foreshock geometry critically depends on the IMF direction.  The standard picture to explain the origin of the ULF foreshock, assumes a stationary IMF and ion beams generated (by different theoretical mechanisms) at the bow shock and backstreaming along magnetic field lines. These beams traveling along stationary field lines are the ultimate cause of ULF waves that, once generated, propagate downstream. Within this general framework, the main purpose of our study is to identify the location of the ULF foreshock boundary under conditions that can be regarded as (at least approximately) stationary. Therefore, assuming that the overall foreshock structure remains in a 
stationary regime, the crossing of the UWFB corresponds to rather mild rotations of the IMF (i.e. small values of $\alpha$). Based on our statistical study of the $\alpha$ angle, we assume that Cluster crosses the quasi-stationary UWFB whenever $\alpha<12.5^\circ$. We choose this limiting angle because we consider that this value is sufficiently small and yet it allows for a substantial number of crossings. On the other hand, relatively large values of $\alpha$ correspond to non-stationary configurations, and therefore we filter these cases out, since the boundary in these cases is also non-stationary.

For $\theta_{\text{Bx}}=45^\circ$ the boundary forms an angle of 87$^\circ \pm 6^\circ$ with respect to the $\hat{\textbf{x}}_{\text{GSE}}$. The observed UWFB is located downstream with respect to the predicted theoretical boundary ($78^\circ$), in agreement with the theoretical prediction \citep{S1988}. We speculate that this difference might be due to the fact that \citet{S1988} criterion is based on the compressive component of the fluctuations, where the amplitude of the waves may be smaller than the ones that we observe. 

Throughout the UWFB a transition take place from FAB distributions (without the presence of ULF waves) to gyrating ion distributions (with the occurrence of ULF waves). However, if we consider the \citet{BM1981a} classification, our statistical boundary is compatible with the presence of diffuse distributions in the downstream region. We emphasize that the large spread and lack of consideration of the IMF cone in the classification performed by \citet{BM1981a} stem from the fact that these determinations are contaminated from several ions distributions. On other hand, our statistical results are in agreement with the UWFB reported by GB86 and the ion foreshock boundary presented by MD98. However, we note that the correlation between waves and the presence of some type of ion distributions is strongly dependent on the mechanism of generation of the ULF waves. Therefore, it is necessary to pursue a detailed investigation of the ion distribution function at both sides of the UWFB to infer any kind of 
correlation between ion distributions and waves. \\

To understand the variation of the boundary with the cone angle $\theta_{Bx}$, we examined the speed of ions propagating along this boundary and compared the obtained results with the classical mechanisms. We have found that the specularly reflected ions are excluded in providing the necessary energy for the wave excitation. Moreover, the hypothesis of adiabatic reflection predicts ion speeds that are larger than those associated with the UWFB. One possible explanation for this behavior is suggested: that the difference may be due to momentum exchange between the incident solar- wind population and the backstreaming particles through the wave-particle interaction resulting from a beam-plasma instability.

A comprehensive understanding of the UWFB in the context of the wave-particle interaction requires the detailed study of the ion distribution function of each for the 102 crossings at both sides of the UWFB boundary to conclude which type of ion distributions are present. This will be the scope of a future work.

\begin{acknowledgments}

NA, DG and CB would like to acknowledge support from grants UBACyT 20020100100315, and PICT 2011-0454. 

NA, CM, DG and CB work was supported by the ECOS-MINCyT cooperation program.

NA acknowledges E. Penou for his assistance.

The authors acknowledge Cluster Active Archive ({http://caa.estec.esa.int/caa/home.xml}) and CDAWeb ({http://cdaweb.gsfc.nasa.gov/}) for the support/distribution of the data used in the present work.

\end{acknowledgments}

\section*{References}


\bibliographystyle{agufull08}

\end{document}